\documentclass[aps,prl,twocolumn,showpacs,superscriptaddress]{revtex4}

\usepackage{amssymb,amsmath}
\usepackage{graphicx}
\usepackage[amssymb,pstricks]{SIunits}
\usepackage{color}

\begin{document}

\title{Confining Light in Deep Subwavelength Electromagnetic Cavities}

\author{V.\ Ginis}
\affiliation{Dept.\ of Applied Physics and Photonics, Vrije Universiteit Brussel,
             Pleinlaan 2, B-1050 Brussel, Belgium}
\author{P.\ Tassin}
\affiliation{Dept.\ of Applied Physics and Photonics, Vrije Universiteit Brussel,
             Pleinlaan 2, B-1050 Brussel, Belgium}
\affiliation{Ames Laboratory-U.S. DOE, and Dept.\ of Physics and Astronomy, Iowa State University, Ames, Iowa 50011, USA}
\author{C.\ M.\ Soukoulis}
\affiliation{Ames Laboratory-U.S. DOE, and Dept.\ of Physics and Astronomy, Iowa State University, Ames, Iowa 50011, USA}
\affiliation{Institute of Electronic Structure and Lasers (IESL), FORTH,\\
             and Dept.\ of Material Science and Technology, University of Crete, 71110 Heraklion, Crete, Greece}
\author{I.\ Veretennicoff}
\affiliation{Dept.\ of Applied Physics and Photonics, Vrije Universiteit Brussel,
             Pleinlaan 2, B-1050 Brussel, Belgium}

\date{July 24, 2010}

\begin{abstract}
We demonstrate that it is possible to confine electromagnetic radiation in cavities that are significantly smaller than the wavelength of the radiation it encapsulates. To this aim, we use the techniques of transformation optics. First, we present a ``perfect cavity'' of arbitrarily small size in which such confined modes can exist.  Furthermore, we show that these eigenmodes have a continuous spectrum and that bending losses are absent, in contrast to what is observed in traditional microcavities. Finally, we introduce an alternative cavity configuration that is less sensitive to material imperfections and still exhibits deep subwavelength modes combined with high quality factor, even if considerable material losses are included. Such a cavity may be interesting for the storage of information in optical data processing and for applications in quantum optics.
\end{abstract}

\pacs{41.20.Jb, 42.70.-a, 42.79.-e}
\maketitle

Transformation optics has recently shed new light on the interaction between electromagnetic radiation and matter~\cite{Ward-1996, Pendry-2006,Leonhardt-2006}. It provides a recipe to design components that guide electromagnetic waves along predetermined curved coordinate lines. The advantage of this technique is that it allows to approach an electromagnetic problem from a geometric perspective, by bending and squeezing the coordinate lines. These geometric distortions of space can then be converted into a medium with well-defined constitutive parameters~\cite{Pendry-2006,Leonhardt-2009}. 
Based on an early idea of Pendry~\cite{Ward-1996}, transformation optics was first used to design a spherical perfect lens~\cite{Pendry-2003}. 
The most exciting example of transformation optics is the invisibility cloak, but it has also been applied for beam manipulation, lenses and  {illusion devices}~\cite{Pendry-2006,Leonhardt-2006,Schurig-2006,Cai-2007,Pendry-2008,Liu-2009,Valentine-2009,Chen-2007,Rahm-2008,Lai-2009,Lai2-2009}. The cloaking idea can moreover be used, e.g., in acoustics, to hide  structures from acoustic waves~\cite{Cummer-2008}, hydrodynamics, to protect coastlines or platforms from tidal waves~\cite{Farhat-2008}, or quantum mechanics for cloaking of matter waves~\cite{Zhang-2008}.

So far, the main focus of transformation optics has been on cloaking and beam manipulation. Here we want to show that these ideas can also be used to design devices that are able to confine electromagnetic energy. Nowadays, this can be achieved with microcavities---the most important implementations being Fabry-Perot, dielectric, and photonic crystal cavities~\cite{Vahala-2003,Noda-2007}---and by the use of electromagnetically induced transparency to slow down or even stop light~\cite{Hau-1999,Fleischhauer-2000}. The characteristics of a microcavity are determined by two important parameters: the quality factor $Q$, which describes the temporal confinement of the electromagnetic field, and the mode volume $V$, which is a measure of its spatial extent~\cite{Vahala-2003}. Indeed, several applications involving optical storage require electromagnetic energy to be confined in a small volume over a long period of time~\cite{Vahala-2003}. Unfortunately, traditional cavities are severely limited in size due to the wavelike nature of light, which imposes a lower limit on the mode volume and hence prevents the miniaturization of photonic components below the wavelength~\cite{Miller-2007}. In addition, the electromagnetic storage systems mentioned above all suffer from fundamental losses, e.g., whispering gallery losses in dielectric microcavities~\cite{Vahala-2003}.

In this letter, we want to present a dielectric cavity of deep subwavelength dimensions in combination with a high quality factor. {We want to stress that we consider here a resonant cavity and not a waveguide structure which always has free propagation---and hence no confinement---in one spatial direction.} We start by recalling the transformation-optical machinery leading to the invisibility cloak. The values of the permittivity and permeability that implement a distortion of the electromagnetic space can be calculated by properly designing the transformation of the coordinate lines~\cite{Pendry-2006,Leonhardt-2009}. In the case of an invisibility cloak, the electromagnetic fields cannot propagate inside the cloaked region, e.g., a sphere with radius $R_1$. This is realized by mapping the physical radius $R_1$ on the origin of the electromagnetic space. Additionally, the outer boundary at radius $R_2$ in physical space is mapped onto itself in electromagnetic space, ensuring a smooth transition into the transformation medium and eliminating reflections. Any continuous coordinate transformation $r' = f(r)$ that satisfies the boundary conditions $f(R_1) = 0$ and $f(R_2) = R_2$ will thus implement the effect of an invisibility cloak.

For a cavity that encapsulates electromagnetic energy, we have to achieve the inverse of a cloak: light rays may not escape from the outer boundary for a cavity. We thus have to ensure that electromagnetic waves cannot pass beyond the outer radius $R_2$. 
Adapting the constraints used for the invisibility cloak, we can impose the boundary condition $f(R_2) = 0$ at the outer boundary and allow the radiation to penetrate into the inner region with radius $R_1$. This amounts to imposing that the transformation function is continuous at this boundary. We therefore need the following boundary conditions for the transformation function of a cavity:
\begin{equation}
f(R_1)= R_1, \;\;\; f(R_2)= 0. 
\label{Eq:TransformationBoundary}
\end{equation}
One can interpret this cavity as a medium that cloaks away the surrounding space, instead of the surrounded space. 
Let us first consider the cylindrical case as shown in Fig.~1.
This setup consists of three regions: $\mathrm{I}$ and $\mathrm{III}$ are vacuum, whereas region $\mathrm{II}$ contains a transformation medium, where we will use a radial coordinate transformation mapping the coordinates $(\rho,\phi,z)$  onto the coordinates $(\rho',\phi',z')$ as defined by the transformation
\begin{equation}
\rho' = \frac{R_1}{R_1-R_2}(\rho-R_2), 
\label{Eq:function}
\end{equation}
while the other coordinates ($\phi,z$) remain unchanged. This coordinate transformation, shown in Fig.~1, is one possible transformation that satisfies the boundary conditions derived above. As the radial coordinate gets folded, the corresponding coordinate lines in Cartesian coordinates follow closed loops: a particle, identified by a positive (negative) $x$-coordinate and moving along a vertical coordinate line in region $(\mathrm{I})$, is bent towards the right (left) in the transformation medium and returns into the vacuum region on the same vertical line, staying bounded in this region for an infinite time. 
\begin{figure}
  \begin{center}
    \includegraphics[clip]{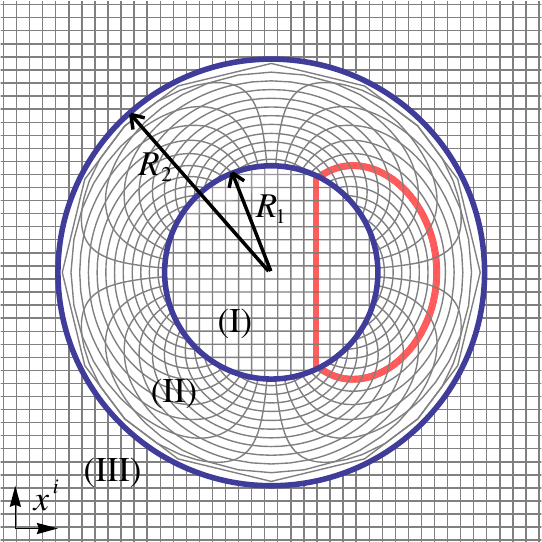}
  \end{center}
  \caption{The coordinate transformation of the perfect cavity. The surrounding space is made invisible through a radial coordinate transformation that maps $R_2$ on the origin in electromagnetic space and is matched with vacuum at $R_1$.  Expressed in Cartesian coordinates $(x^i)$, the coordinate lines get folded back on themselves in closed curves. This is the origin of the perfect confinement.}
  \label{Fig:PerfectCavityTransformationFunction}
\end{figure}

In order to confirm this geometrical picture, we have calculated the bounded modes of this system. Since the cavity is a linear system with cylindrical symmetry, the modes can be written as
\begin{equation}
\mathbf{E}(\mathbf{r},t) = \mathbf{E}(\rho)\mathrm{e}^{\mathrm{i}(m\phi-\omega t)},
\end{equation}
where $m$ is an integer quantifying the angular momentum of the mode and $\omega$ is an eigenfrequency. There is no $z$-dependence of the solutions, since we are considering here an infinite cylinder. In general, $\omega$ can be a complex value $\omega = \omega'+\mathrm{i}\omega''$, where the imaginary part is related to the losses of the electromagnetic energy due to dissipation or radiation. The eigenfrequencies and their corresponding eigenmodes are determined by calculating the solutions of Maxwell's equations in each region ($\mathrm{I}, \mathrm{II}, \mathrm{III}$) and combining them with the proper boundary conditions.
Without loss of generality we can assume {TM polarization}. In the vacuum regions ($\mathrm{I}$) and ($\mathrm{III}$), Maxwell's equations can be combined to Helmholtz' equation.
One can derive that the magnetic field in these regions $(\mathrm{I})$ and $(\mathrm{III})$ is given by:
\begin{eqnarray}
\label{Eq:EInMicroCavity}
H^z_\mathrm{I}(\rho) &=& A\ J_{m}(k_0\rho) + B\ Y_{m}(k_0\rho),\\
\label{Eq:EOutMicroCavity}
H^z_\mathrm{III}(\rho) &=& E\ H_{m}^{(1)}(k_0\rho) + F\ H_{m}^{(2)}(k_0\rho),
\end{eqnarray}
where $k_0 = \omega/c$, $J_m$ and $Y_m$ are the Bessel functions of the first and second kind, $H_m^{(1)}$ and $H_m^{(2)}$ are the Hankel functions of first and second kind, and $(A,B,E,F)$ are complex integration constants; we may set $B = F = 0$ to impose finite energy and Sommerfeld's radiation condition.
Inside the transformation medium (region $\mathrm{II}$), our analysis has shown that the magnetic fields have a similar form with the radial coordinate replaced by $f(\rho)$: 
\begin{equation}
\label{Eq:EInsideCloak}
H^z_{\mathrm{II}}(\rho) = C\ J_m(k_0f(\rho))+ D\ Y_m(k_0f(\rho)),
\end{equation}
where, once again, ($C,D$) are arbitrary complex numbers. At the interface between two materials, the tangential components of the electric and magnetic fields must be continuous. When we apply these conditions at $\rho = R_1$ and $\rho = R_2$, we find a set of four equations: 
\begin{align}
\label{Eq:DispersionRelationBegin}
&A\ J_m(k_0R_1) = C\ J_m(k_0f(R_1))+D\ Y_m(k_0f(R_1)),\\
\label{Eq:DispersionRelation2}
&A\ J'_m(k_0R_1) = C\ \frac{f(R_1)}{R_1}J'_m(k_0f(R_1))\nonumber\\&\qquad\qquad+ D\ \frac{f(R_1)}{R_1}Y'_m(k_0f(R_1)),\\
\label{Eq:DispersionRelation3}
&C\ J_m(k_0f(R_2))+D\ Y_m(k_0f(R_2)) = E\ H^{(1)}_m(k_0R_2),\\
\label{Eq:DispersionRelationEnd}
&C\ \frac{f(R_2)}{R_2}J'_m(k_0f(R_2))+ D\ \frac{f(R_2)}{R_2}Y'_m(k_0f(R_2))\nonumber\\ &\qquad\qquad= E\ H'^{(1)}_m(k_0R_2),
\end{align}
where the prime $(')$ denotes differentiation with respect to the radial coordinate $\rho$.

Surprisingly, when we apply the transformation function defined by Eqs.~(\ref{Eq:TransformationBoundary}), we notice that there is no quantization of the eigenfrequencies. This means that modes with an arbitrary value of $\omega'$ can exist inside this cavity, even if the free-space wavelength ($\lambda_0 = 2\pi c/\omega'$) is many times larger than the dimensions of the cavity. In Fig.~2, we show the magnetic field of such a deep subwavelength mode in which the free space wavelength is three hundred times larger than the outer radius of the cavity. From this figure, one can observe that the field is exactly zero in the outside region ($\rho>R_2$). This means that the electromagnetic energy is entirely located inside the cavity: these subwavelength modes are thus characterized by an infinite quality factor. One might therefore call this device a ``perfect cavity.'' Since the boundary conditions (\ref{Eq:DispersionRelationBegin})-(\ref{Eq:DispersionRelationEnd}) are independent of $R_1$, the perfect cavity retains its properties for arbitrary values of the radius of the inner vacuum region.
\begin{figure}
  \centering
  \includegraphics[clip]{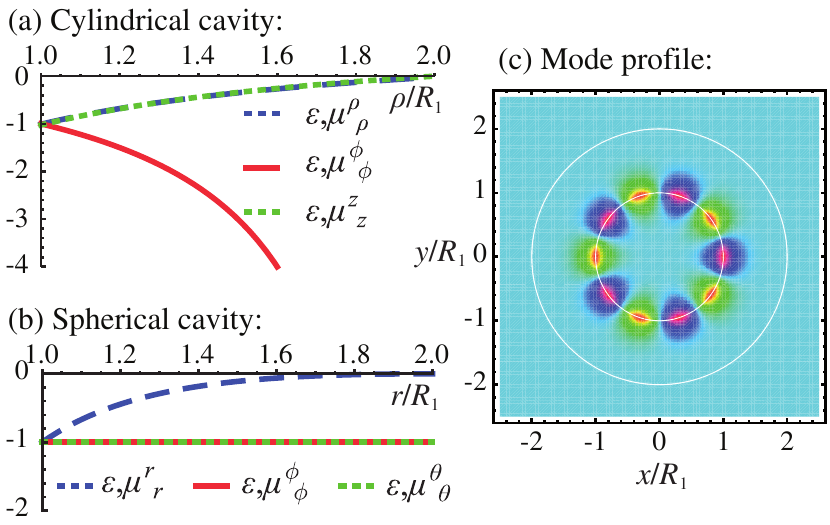}
  \caption{{(a) Material parameters of the perfect cavity as defined by Eq.~(\ref{Eq:function}). (b) The parameters of the equivalent spherical implementation. (c) The magnetic field distribution of a perfectly confined cavity mode with $R_2/\lambda_0 = 0.0032$; the magnetic field is exactly zero in the outer region, implying the absence of energy radiated away to infinity.}}
  \label{Fig:PerfectCavityDensityPlot}
\end{figure}

The values of the permittivity and permeability required to materialize this cavity can be determined using the equivalence relations of transformation optics~\cite{Pendry-2006,Leonhardt-2009}. With the transformation function as given in Eq.~(\ref{Eq:function}), we find the following nontrivial components:
\begin{align}
\epsilon^{\rho}_{\phantom{\rho}\rho} &= \mu^{\rho}_{\phantom{\rho}\rho} = \frac{\rho-R_2}{\rho},\;\;\;
\epsilon^{\phi}_{\phantom{\phi}\phi} = \mu^{\phi}_{\phantom{\phi}\phi} = \frac{\rho}{\rho-R_2},\\
\epsilon^{z}_{\phantom{z}z} &= \mu^{z}_{\phantom{z}z} = \frac{R_1^2}{(R_1-R_2)^2}\frac{\rho-R_2}{\rho}.
\end{align}
The variation of these components as a function of the physical coordinate $\rho$ is shown in Fig.~2(a)-(b).
Each component of $\epsilon^i_j$ and $\mu^i_j$ has a negative value, imposing the use of left-handed materials. It is easily seen that any transformation medium satisfying $f(R_1)=R_1$ and $f(R_2)=0$ will have a region with left-handed materials. Subsequently, we notice the behavior at the outer boundary, which is analogous to the inner boundary of the invisibility cloaks: the radial component becomes zero, while the angular component tends to minus infinity. We also studied a spherical implementation of this cavity and we found that it also exhibits a continuum of deep subwavelength modes with perfect quality factor. The material implementation is more realistic than the cylindrical case as it does not require any component of the constitutive parameters that tend to infinity, as shown in Fig.~2(b).

\begin{figure}
  \centering
   \includegraphics[clip]{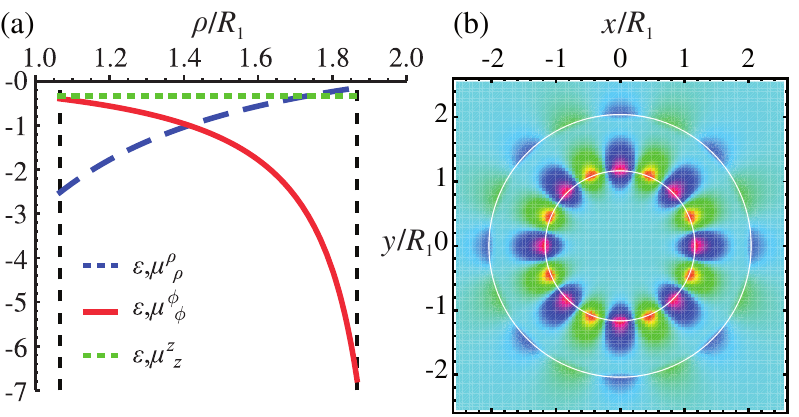}
  \caption{{The non-singular cavity is constructed from a perfect cavity, but with thin rims $\Delta R_1$ and $\Delta R_2$ removed at the inner and outer boundaries (vertical broken lines). (a) The resulting material parameters do not assume extreme values. (b) The magnetic field distribution inside this cavity, corresponding to a deep subwavelength solution with $R_\mathrm{out}/\lambda_0 = 0.19$. There is a small part of the mode situated outside the outer radius, corresponding to a quality factor $Q = 1.1\times10^{10}$.}}
  \label{Fig:PerfectCavityFieldExample}
\end{figure}

Unfortunately, we have found that the design discussed above is highly sensitive to the value of the material parameters. Essentially being a strange kind of cloak, one might expect the same kind of sensitivity: as material parameters deviate from the ideal values, an invisibility cloak retains its cloaking characteristics, albeit less performant~\cite{Schurig-2006}. In this case, however, when we perturb the cavity by taking away a little rim from the outer boundary, we notice that the eigenmodes disappear completely. In the second part of this letter, we therefore propose an alternative design of the cavity that eliminates this singularity. When the outer boundary is not perfectly mapped onto the origin in electromagnetic space, our simulations show that energy is radiated away to infinity. This prohibits the existence of confined modes. We can reintroduce deep subwavelength modes with an additional perturbation at the inner boundary, thus removing the impedance matching. This will lead under certain conditions to destructive interference of the outside field. This idea is supported by the solutions of the dispersion relation and by our full-wave simulations with a finite-element solver (COMSOL Multiphysics). We start from a perfect cavity with the transformation function
\begin{equation}
\rho' = \frac{R_1}{\sqrt{R_2^2-R_1^2}}\sqrt{R_2^2-\rho^2},
\label{Eq:transformation2}
\end{equation} 
from which we now slice off thin rims at both the inner and outer boundaries, i.e., the material is situated between the radii $R_1+\Delta R_1$ and $R_2-\Delta R_2$. The material parameters that constitute this transformation are shown in Fig.~3(a), where we have chosen perturbations of a few percent; this reduces the constraints on the materials significantly (the permittivity and permeability range from $-0.15$ to $-6.76$). For this configuration, the dispersion relation allows for one single mode solution, for every integer value of the angular momentum parameter $m$. 
In Fig.~3(b), we plot the magnetic field of such a mode with $m=8$, which has $R_2/\lambda_0 = 0.19$, i.e., the wavelength is more than five times larger than the outer radius of the cavity, and a quality factor of $1.1\times10^{10}$. We notice that it has the same structure as for the perfect cavity. 

\begin{figure}
  \centering
  \includegraphics[clip]{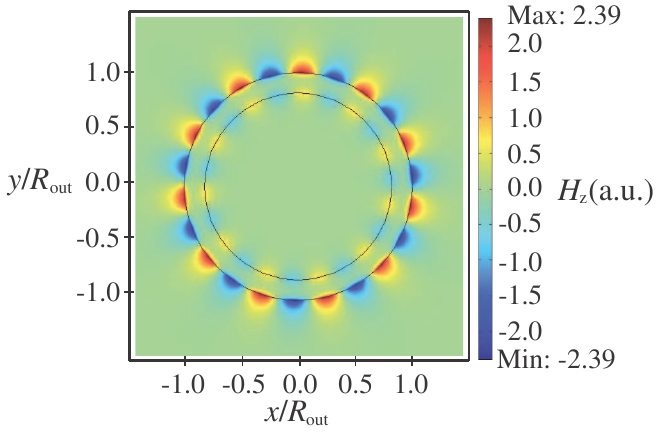}
  \caption{The magnetic field of an eigenmode of the cavity including material losses. When we fix the geometric parameters at $R_1 = 0.0015R_2$, $\Delta R_1=0.75R_2$, $\Delta R_2 = 0.083R_2$, $\alpha=1.0$, $\beta=1.0\times 10^{-2}$, and $\gamma=1.5825\times 10^{-2}$, we find a subwavelength solution with angular mode number $m=11$ at $R_\mathrm{out} / \lambda=0.14$ with $Q=1983$. The real parts of $\epsilon$ and $\mu$ stay bounded between $0$ and $-5.2$.}
  \label{Fig:F5_Comsol}
\end{figure}

{
The reader might object that the typical high losses in the metamaterials required for the cavity's implementation will adversely affect the subwavelength modes.  We have therefore extended the formalism to calculate the eigenfrequencies of the cavity made from lossy metamaterials. By considering TM polarization, only one permeability ($\mu^z_{\phantom{z}z}$) and two permittivities ($\epsilon^\rho_{\phantom{\rho}\rho}$, $\epsilon^\phi_{\phantom{\phi}\phi}$) need to be implemented. We now introduce the loss tangents $\alpha$ and $\beta$ of the permeability and the permittivity, respectively. We can show that the effect of these loss tangents boils down to replacing the transformation $f(\rho)$ by $g(\rho)=\sqrt{(1+\mathrm{i}\alpha)(1+\mathrm{i}\beta)}f(\rho)$, and to add a factor $(1+\mathrm{i}\beta)^{-1}$ to the right-hand side of Eqs.~(\ref{Eq:DispersionRelation2})-(\ref{Eq:DispersionRelationEnd}). We have determined the resulting quality factor by using realistic loss tangents that are found for fishnet structures operating at optical frequencies~\cite{Dolling-2007}, {and that have been shown to converge well to bulk values for multiple-layer fishnets that we need here.} The negative permeability, which occurs close to the magnetic resonance of the fishnet, has typically a large loss tangent of approximately $\alpha=1$, whereas the permittivities have a nonresonant Drude dispersion with typical loss tangent $\beta=10^{-2}$. Our results show that such a lossy cavity still supports subwavelength modes, although with much lower quality factor. We have found that the quality factor can be increased by improving the impedance matching at the inner boundary by filling the inner region with a nonmagnetic material with loss tangent $\gamma \approx \beta$. This procedure led us after optimization with COMSOL Multiphysics to the results shown in Fig.~\ref{Fig:F5_Comsol}: we find again deep subwavelength eigenmodes ($R_\mathrm{out}/\lambda=0.14$) with quality factors up to $Q \approx 2000$. These Q-factors are higher than the theoretical upper limit predicted for plasmonic cavities~\cite{Min-2009}.
}

In this letter, we have demonstrated that electromagnetic radiation can be confined in cavities with dimensions well below the wavelength of the radiation. {These cavities do not require materials with extremely high indices of refraction and therefore do not suffer from a breakdown of the effective medium approximation that would otherwise arise due to the discrete atomic nature of all real materials.}

We thank Ingo Fischer for inspiring conversations on using invisibility cloaks as electromagnetic cavities. Work at the VUB was supported by  BelSPO (Grant No.\ IAP6/10 Photonics@be), the FWO-Vlaanderen, and the Research Council (OZR) of the VUB. Work at Ames Laboratory was supported by the Department of Energy (Basic Energy Sciences) under Contract No.\ DE-AC02-07CH11358. P.\ T.\ acknowledges the FWO-Vlaanderen and the Belg.\ Am.\ Educ.\ Found.\ for financial support.


\begin{thebibliography}{22}
\expandafter\ifx\csname natexlab\endcsname\relax\def\natexlab#1{#1}\fi
\expandafter\ifx\csname bibnamefont\endcsname\relax
  \def\bibnamefont#1{#1}\fi
\expandafter\ifx\csname bibfnamefont\endcsname\relax
  \def\bibfnamefont#1{#1}\fi
\expandafter\ifx\csname citenamefont\endcsname\relax
  \def\citenamefont#1{#1}\fi
\expandafter\ifx\csname url\endcsname\relax
  \def\url#1{\texttt{#1}}\fi
\expandafter\ifx\csname urlprefix\endcsname\relax\def\urlprefix{URL }\fi
\providecommand{\bibinfo}[2]{#2}
\providecommand{\eprint}[2][]{\url{#2}}

\bibitem[{\citenamefont{{Ward} and {Pendry}}(1996)}]{Ward-1996}
\bibinfo{author}{\bibfnamefont{A.~J.} \bibnamefont{{Ward}}} \bibnamefont{and}
  \bibinfo{author}{\bibfnamefont{J.~B.} \bibnamefont{{Pendry}}},
  \bibinfo{journal}{J.\ Mod.\ Phys.} \textbf{\bibinfo{volume}{43}},
  \bibinfo{pages}{773} (\bibinfo{year}{1996}).

\bibitem[{\citenamefont{{Pendry} et~al.}(2006)\citenamefont{{Pendry},
  {Schurig}, and {Smith}}}]{Pendry-2006}
\bibinfo{author}{\bibfnamefont{J.~B.} \bibnamefont{{Pendry}}},
  \bibinfo{author}{\bibfnamefont{D.}~\bibnamefont{{Schurig}}},
  \bibnamefont{and} \bibinfo{author}{\bibfnamefont{D.~R.}
  \bibnamefont{{Smith}}}, \bibinfo{journal}{Science}
  \textbf{\bibinfo{volume}{312}}, \bibinfo{pages}{1780} (\bibinfo{year}{2006}).

\bibitem[{\citenamefont{{Leonhardt}}(2006)}]{Leonhardt-2006}
\bibinfo{author}{\bibfnamefont{U.}~\bibnamefont{{Leonhardt}}},
  \bibinfo{journal}{Science} \textbf{\bibinfo{volume}{312}},
  \bibinfo{pages}{1777} (\bibinfo{year}{2006}).

\bibitem[{\citenamefont{{Leonhardt} and {Philbin}}(2009)}]{Leonhardt-2009}
\bibinfo{author}{\bibfnamefont{U.}~\bibnamefont{{Leonhardt}}} \bibnamefont{and}
  \bibinfo{author}{\bibfnamefont{T.~G.} \bibnamefont{{Philbin}}},
  \bibinfo{journal}{Prog.\ Opt.} \textbf{\bibinfo{volume}{53}},
  \bibinfo{pages}{70} (\bibinfo{year}{2009}).

\bibitem[{\citenamefont{{Pendry} and {Ramakrishna}}(2003)}]{Pendry-2003}
\bibinfo{author}{\bibfnamefont{J.~B.} \bibnamefont{{Pendry}}} \bibnamefont{and}
  \bibinfo{author}{\bibfnamefont{S.~A.} \bibnamefont{{Ramakrishna}}},
  \bibinfo{journal}{J.\ Phys.\ Cond.\ Matter} \textbf{\bibinfo{volume}{15}},
  \bibinfo{pages}{6345} (\bibinfo{year}{2003}).

\bibitem[{\citenamefont{{Schurig} et~al.}(2006)\citenamefont{{Schurig},
  {Pendry}, and {Smith}}}]{Schurig-2006}
\bibinfo{author}{\bibfnamefont{D.}~\bibnamefont{{Schurig}}},
  \bibinfo{author}{\bibfnamefont{J.~B.} \bibnamefont{{Pendry}}},
  \bibnamefont{and} \bibinfo{author}{\bibfnamefont{D.~R.}
  \bibnamefont{{Smith}}}, \bibinfo{journal}{Optics Express}
  \textbf{\bibinfo{volume}{14}}, \bibinfo{pages}{9794} (\bibinfo{year}{2006}).

\bibitem[{\citenamefont{{Cai} et~al.}(2007)\citenamefont{{Cai}, {Chettiar},
  {Kildishev}, and {Shalaev}}}]{Cai-2007}
\bibinfo{author}{\bibfnamefont{W.}~\bibnamefont{{Cai}}},
  \bibinfo{author}{\bibfnamefont{U.~K.} \bibnamefont{{Chettiar}}},
  \bibinfo{author}{\bibfnamefont{A.~V.} \bibnamefont{{Kildishev}}},
  \bibnamefont{and} \bibinfo{author}{\bibfnamefont{V.~M.}
  \bibnamefont{{Shalaev}}}, \bibinfo{journal}{Nature Photon.}
  \textbf{\bibinfo{volume}{1}}, \bibinfo{pages}{224} (\bibinfo{year}{2007}).

\bibitem[{\citenamefont{{Li} and {Pendry}}(2008)}]{Pendry-2008}
\bibinfo{author}{\bibfnamefont{J.}~\bibnamefont{{Li}}} \bibnamefont{and}
  \bibinfo{author}{\bibfnamefont{J.~B.} \bibnamefont{{Pendry}}},
  \bibinfo{journal}{Phys.\ Rev.\ Lett.} \textbf{\bibinfo{volume}{101}},
  \bibinfo{pages}{203901} (\bibinfo{year}{2008}).

\bibitem[{\citenamefont{{Liu} et~al.}(2009)\citenamefont{{Liu}, {Ji}, {Mock},
  {Chin}, {Cui}, and {Smith}}}]{Liu-2009}
\bibinfo{author}{\bibfnamefont{R.}~\bibnamefont{{Liu}}},
  \bibinfo{author}{\bibfnamefont{C.}~\bibnamefont{{Ji}}},
  \bibinfo{author}{\bibfnamefont{J.~J.} \bibnamefont{{Mock}}},
  \bibinfo{author}{\bibfnamefont{J.~Y.} \bibnamefont{{Chin}}},
  \bibinfo{author}{\bibfnamefont{T.~J.} \bibnamefont{{Cui}}}, \bibnamefont{and}
  \bibinfo{author}{\bibfnamefont{D.~R.} \bibnamefont{{Smith}}},
  \bibinfo{journal}{Science} \textbf{\bibinfo{volume}{323}},
  \bibinfo{pages}{366} (\bibinfo{year}{2009}).

\bibitem[{\citenamefont{{Valentine} et~al.}(2009)\citenamefont{{Valentine},
  {Li}, {Zentgraf}, {Bartal}, and {Zhang}}}]{Valentine-2009}
\bibinfo{author}{\bibfnamefont{J.}~\bibnamefont{{Valentine}}},
  \bibinfo{author}{\bibfnamefont{J.}~\bibnamefont{{Li}}},
  \bibinfo{author}{\bibfnamefont{T.}~\bibnamefont{{Zentgraf}}},
  \bibinfo{author}{\bibfnamefont{G.}~\bibnamefont{{Bartal}}}, \bibnamefont{and}
  \bibinfo{author}{\bibfnamefont{X.}~\bibnamefont{{Zhang}}},
  \bibinfo{journal}{Nature Mater.} \textbf{\bibinfo{volume}{8}},
  \bibinfo{pages}{568} (\bibinfo{year}{2009}).

\bibitem[{\citenamefont{{Chen} and {Chan}}(2007)}]{Chen-2007}
\bibinfo{author}{\bibfnamefont{H.}~\bibnamefont{{Chen}}} \bibnamefont{and}
  \bibinfo{author}{\bibfnamefont{C.~T.} \bibnamefont{{Chan}}},
  \bibinfo{journal}{Appl.\ Phys.\ Lett.} \textbf{\bibinfo{volume}{90}},
  \bibinfo{pages}{241105} (\bibinfo{year}{2007}).

\bibitem[{\citenamefont{{Rahm} et~al.}(2008)\citenamefont{{Rahm}, {Schurig},
  {Roberts}, {Cummer}, {Smith}, and {Pendry}}}]{Rahm-2008}
\bibinfo{author}{\bibfnamefont{M.}~\bibnamefont{{Rahm}}},
  \bibinfo{author}{\bibfnamefont{D.}~\bibnamefont{{Schurig}}},
  \bibinfo{author}{\bibfnamefont{D.~A.} \bibnamefont{{Roberts}}},
  \bibinfo{author}{\bibfnamefont{S.~A.} \bibnamefont{{Cummer}}},
  \bibinfo{author}{\bibfnamefont{D.~R.} \bibnamefont{{Smith}}},
  \bibnamefont{and} \bibinfo{author}{\bibfnamefont{J.~B.}
  \bibnamefont{{Pendry}}}, \bibinfo{journal}{Photon.~Nanostruct.: Fundam.
  Applic.} \textbf{\bibinfo{volume}{6}}, \bibinfo{pages}{87}
  (\bibinfo{year}{2008}).

\bibitem[{\citenamefont{{Lai}}(2009)}]{Lai-2009}
\bibinfo{author}{\bibfnamefont{Y.}~\bibnamefont{{Lai}}},
  \bibinfo{author}{\bibfnamefont{H.}~\bibnamefont{{Chen}}},
  \bibinfo{author}{\bibfnamefont{Z.} \bibnamefont{{Zhang}}},
  \bibnamefont{and}
  \bibinfo{author}{\bibfnamefont{C.~T.} \bibnamefont{{Chan}}},
   \bibinfo{journal}{Phys.\ Rev.\ Lett.} \textbf{\bibinfo{volume}{102}}, \bibinfo{pages}{093901}
  (\bibinfo{year}{2009}).

\bibitem[{\citenamefont{{Lai2}}(2009)}]{Lai2-2009}
\bibinfo{author}{\bibfnamefont{Y.}~\bibnamefont{{Lai}}},
  \bibinfo{author}{\bibfnamefont{J.}~\bibnamefont{{Ng}}},
  \bibinfo{author}{\bibfnamefont{H.} \bibnamefont{{Chen}}},
   \bibinfo{author}{\bibfnamefont{D.} \bibnamefont{{Han}}},
  \bibinfo{author}{\bibfnamefont{J.} \bibnamefont{{Xiao}}},
  \bibinfo{author}{\bibfnamefont{Z.} \bibnamefont{{Zhang}}},
  \bibnamefont{and}
  \bibinfo{author}{\bibfnamefont{C.~T.} \bibnamefont{{Chan}}},
   \bibinfo{journal}{Phys.\ Rev.\ Lett.} \textbf{\bibinfo{volume}{102}}, \bibinfo{pages}{253902}
  (\bibinfo{year}{2009}).  

\bibitem[{\citenamefont{{Cummer} et~al.}(2008)\citenamefont{{Cummer}, {Popa},
  {Schurig}, {Smith}, {Pendry}, {Rahm}, and {Starr}}}]{Cummer-2008}
\bibinfo{author}{\bibfnamefont{S.~A.} \bibnamefont{{Cummer}}},
  \bibinfo{author}{\bibfnamefont{B.-I.} \bibnamefont{{Popa}}},
  \bibinfo{author}{\bibfnamefont{D.}~\bibnamefont{{Schurig}}},
  \bibinfo{author}{\bibfnamefont{D.~R.} \bibnamefont{{Smith}}},
  \bibinfo{author}{\bibfnamefont{J.}~\bibnamefont{{Pendry}}},
  \bibinfo{author}{\bibfnamefont{M.}~\bibnamefont{{Rahm}}}, \bibnamefont{and}
  \bibinfo{author}{\bibfnamefont{A.}~\bibnamefont{{Starr}}},
  \bibinfo{journal}{Phys.\ Rev.\ Lett.} \textbf{\bibinfo{volume}{100}},
  \bibinfo{pages}{024301} (\bibinfo{year}{2008}).

\bibitem[{\citenamefont{{Farhat} et~al.}(2008)\citenamefont{{Farhat}, {Enoch},
  {Guenneau}, and {Movchan}}}]{Farhat-2008}
\bibinfo{author}{\bibfnamefont{M.}~\bibnamefont{{Farhat}}},
  \bibinfo{author}{\bibfnamefont{S.}~\bibnamefont{{Enoch}}},
  \bibinfo{author}{\bibfnamefont{S.}~\bibnamefont{{Guenneau}}},
  \bibnamefont{and} \bibinfo{author}{\bibfnamefont{A.~B.}
  \bibnamefont{{Movchan}}}, \bibinfo{journal}{Phys.\ Rev.\ Lett.}
  \textbf{\bibinfo{volume}{101}}, \bibinfo{pages}{134501}
  (\bibinfo{year}{2008}).

\bibitem[{\citenamefont{{Zhang} et~al.}(2008)\citenamefont{{Zhang}, {Genov},
  {Sun}, and {Zhang}}}]{Zhang-2008}
\bibinfo{author}{\bibfnamefont{S.}~\bibnamefont{{Zhang}}},
  \bibinfo{author}{\bibfnamefont{D.~A.} \bibnamefont{{Genov}}},
  \bibinfo{author}{\bibfnamefont{C.}~\bibnamefont{{Sun}}}, \bibnamefont{and}
  \bibinfo{author}{\bibfnamefont{X.}~\bibnamefont{{Zhang}}},
  \bibinfo{journal}{Phys.\ Rev.\ Lett.} \textbf{\bibinfo{volume}{100}},
  \bibinfo{pages}{123002} (\bibinfo{year}{2008}).

\bibitem[{\citenamefont{{Vahala}}(2003)}]{Vahala-2003}
\bibinfo{author}{\bibfnamefont{K.~J.} \bibnamefont{{Vahala}}},
  \bibinfo{journal}{Nature} \textbf{\bibinfo{volume}{424}},
  \bibinfo{pages}{839} (\bibinfo{year}{2003}).

\bibitem[{\citenamefont{{Noda} et~al.}(2007)\citenamefont{{Noda}, {Fujita}, and
  {Asano}}}]{Noda-2007}
\bibinfo{author}{\bibfnamefont{S.}~\bibnamefont{{Noda}}},
  \bibinfo{author}{\bibfnamefont{M.}~\bibnamefont{{Fujita}}}, \bibnamefont{and}
  \bibinfo{author}{\bibfnamefont{T.}~\bibnamefont{{Asano}}},
  \bibinfo{journal}{Nature Photon.} \textbf{\bibinfo{volume}{1}},
  \bibinfo{pages}{449} (\bibinfo{year}{2007}).

\bibitem[{\citenamefont{{Hau} et~al.}(1999)\citenamefont{{Hau}, {Harris},
  {Dutton}, and {Behroozi}}}]{Hau-1999}
\bibinfo{author}{\bibfnamefont{L.~V.} \bibnamefont{{Hau}}},
  \bibinfo{author}{\bibfnamefont{S.~E.} \bibnamefont{{Harris}}},
  \bibinfo{author}{\bibfnamefont{Z.}~\bibnamefont{{Dutton}}}, \bibnamefont{and}
  \bibinfo{author}{\bibfnamefont{C.~H.} \bibnamefont{{Behroozi}}},
  \bibinfo{journal}{Nature} \textbf{\bibinfo{volume}{397}},
  \bibinfo{pages}{594} (\bibinfo{year}{1999}).

\bibitem[{\citenamefont{{Fleischhauer} and {Lukin}}(2000)}]{Fleischhauer-2000}
\bibinfo{author}{\bibfnamefont{M.}~\bibnamefont{{Fleischhauer}}}
  \bibnamefont{and} \bibinfo{author}{\bibfnamefont{M.~D.}
  \bibnamefont{{Lukin}}}, \bibinfo{journal}{Phys.\ Rev.\ Lett.}
  \textbf{\bibinfo{volume}{84}}, \bibinfo{pages}{5094} (\bibinfo{year}{2000}).

\bibitem[{\citenamefont{{Miller}}(2007)}]{Miller-2007}
\bibinfo{author}{\bibfnamefont{D.~A.~B.} \bibnamefont{{Miller}}},
  \bibinfo{journal}{\josab} \textbf{\bibinfo{volume}{24}}, \bibinfo{pages}{A1}
  (\bibinfo{year}{2007}).

\bibitem[{\citenamefont{{Dolling}}(2007)}]{Dolling-2007}
\bibinfo{author}{\bibfnamefont{G.}~\bibnamefont{{Dolling}}},
  \bibinfo{author}{\bibfnamefont{M.}~\bibnamefont{{Wegener}}},
  \bibinfo{author}{\bibfnamefont{C.~M.} \bibnamefont{{Soukoulis}}},
  \bibnamefont{and}
  \bibinfo{author}{\bibfnamefont{S.}~\bibnamefont{{Linden}}},
   \bibinfo{journal}{Optics Express} \textbf{\bibinfo{volume}{18}}, \bibinfo{pages}{11536}
  (\bibinfo{year}{2007}).


\bibitem[{\citenamefont{{Min}}(2009)}]{Min-2009}
\bibinfo{author}{\bibfnamefont{B.}~\bibnamefont{{Min}}},
  \bibinfo{author}{\bibfnamefont{E.}~\bibnamefont{{Ostby}}},
  \bibinfo{author}{\bibfnamefont{V.} \bibnamefont{{Sorger}}},
  \bibinfo{author}{\bibfnamefont{E.}~\bibnamefont{{Ulin-Avila}}},
  \bibinfo{author}{\bibfnamefont{L.} \bibnamefont{{Yang}}},
  \bibinfo{author}{\bibfnamefont{X.}~\bibnamefont{{Zhang}}},
  \bibnamefont{and}
  \bibinfo{author}{\bibfnamefont{K.} \bibnamefont{{Vahala}}},
   \bibinfo{journal}{Nature} \textbf{\bibinfo{volume}{457}}, \bibinfo{pages}{455}
  (\bibinfo{year}{2009}).
  

\end{thebibliography}

\end{document}